\documentclass[conference]{IEEEtran}
\usepackage{graphicx}%
\usepackage{bm}%
\usepackage{amsfonts}
\usepackage{amssymb}
\usepackage{url}
\usepackage{times}
\usepackage{subfigure}
\usepackage{enumitem}
\usepackage{amsmath} 
\usepackage{latexsym}
\usepackage{hhline}
\usepackage{cite}
\usepackage{caption}
\usepackage{algorithm}
\usepackage{algpseudocode}
\usepackage{multirow} 
\usepackage{xcolor}
\usepackage{epstopdf}
\usepackage{aligned-overset}

\allowdisplaybreaks

\begin{document}

\title{Neural Network-Based Delay–Doppler-Assisted Channel Estimation for OFDM}
\author{\IEEEauthorblockN{
Mingcheng Nie\IEEEauthorrefmark{1},
Hao Chang\IEEEauthorrefmark{1},
Shuangyang Li\IEEEauthorrefmark{2},
Haiyao Yu\IEEEauthorrefmark{1},
Jiafu Hao\IEEEauthorrefmark{1},
Yonghui Li\IEEEauthorrefmark{1},
}
\IEEEauthorblockA{
\IEEEauthorrefmark{1}The University of Sydney, Sydney, Australia\\
\IEEEauthorrefmark{2}Technische Universit{\"a}t Berlin, Berlin, Germany\\
\emph{(Invited Paper)}
}}

\markboth{Journal of \LaTeX\ Class Files,~Vol.~14, No.~8, August~2021}%
{Shell \MakeLowercase{\textit{et al.}}: A Sample Article Using IEEEtran.cls for IEEE Journals}


\maketitle

\begin{abstract}

Conventional orthogonal frequency division multiplexing (OFDM) channel estimation relies on single-tap estimation and time-frequency (TF) interpolation, which becomes unreliable in high-mobility channels because Doppler-induced inter-carrier interference (ICI) invalidates the underlying element-wise TF model. This paper proposes a neural-network-based delay-Doppler (DD)-assisted channel estimation framework for OFDM over doubly selective channels. We first derive an ICI-aware TF domain input-output relation and formulate channel estimation as a DD recovery problem. Unlike conventional sparse recovery approaches, the proposed framework does not require the equivalent DD domain channel vector to be strictly sparse, thereby accommodating the leakage induced by fractional delay and Doppler shifts. Since the data symbols are unknown during channel estimation, the sensing matrix is constructed using only the known pilot symbols. As a result, data-induced interference is not explicitly modeled, leading to a structured mismatch in the pilot observations. To tackle this challenge, the adopted network iteratively exchanges observation- and channel-domain features through the sensing matrix to learn the mapping from these contaminated observations to the equivalent DD domain channel, which is subsequently used to reconstruct the TF-domain channel. Simulation results show that the proposed method achieves lower normalized mean-square error and bit-error rate than conventional OFDM estimators.

\end{abstract}

\section{Introduction}

Orthogonal frequency division multiplexing (OFDM) has been widely adopted in modern wireless communication systems due to its robustness against frequency-selective fading and low-complexity implementation~\cite{3GPP38211}. With cyclic prefix (CP) insertion and removal, a frequency-selective channel can be decomposed into approximately frequency-flat parallel subchannels, allowing simple single-tap channel estimation (STE) and equalization.
However, this advantage relies on the preservation of perfect subcarrier orthogonality, which becomes overly idealistic in high-mobility scenarios, particularly at high carrier frequency~\cite{wei2021orthogonal,nie2023improving}. In such environments, Doppler-induced non-uniform frequency shifts destroy subcarrier orthogonality and cause severe inter-carrier interference (ICI). Meanwhile, the reduced channel coherence time leads to rapid channel variations across OFDM symbols, thereby requiring more frequent channel estimation \cite{nie2024uplink}. As practical systems must constrain pilot overhead to maintain spectral efficiency, conventional time-frequency (TF) domain channel estimation methods may experience significant performance degradation in highly dynamic channels.

Recently, delay-Doppler (DD) domain communications and related waveforms \cite{wong2026fairness}, such as orthogonal time-frequency space (OTFS)~\cite{hadani2017orthogonal,11373535,10654761,zhang2025deep}, have attracted significant attention for enabling reliable communication over high-mobility channels. By representing a rapidly time-varying TF domain channel in the DD domain, the channel can often be described in a more compact, quasi-static, and sparse form. This property provides a fresh perspective on the design of channel estimation. For instance, the embedded-pilot scheme in~\cite{raviteja2019embedded} places a high-power pilot symbol in the DD domain with sufficient guard space. Owing to the convolutional DD domain input-output relation and the quasi-static nature of the DD domain channel, the channel parameters can be estimated by comparing the received and transmitted DD domain pilot symbols. Thus, the resulting symbol detection performance is much better than in OFDM systems, especially in high-mobility environments \cite{10225901,10683123,11153399}.

Nevertheless, such DD domain pilot designs are not directly compatible with practical OFDM frame structures. In particular, an embedded DD domain pilot may spread over the entire TF domain, causing considerable interference to OFDM data symbols. 
This motivates an important question: can the benefits of DD domain channel representation be exploited while retaining conventional TF domain OFDM pilots? Answering this question would enable improved channel estimation in OFDM systems without requiring dense TF pilot patterns or resorting to DD waveforms. Nevertheless, only limited studies have explored this direction. In~\cite{hu2024cross}, a cross-domain channel estimation (CDCE) method was proposed for OTFS transmission with TF domain pilots, where initial TF domain channel estimates are obtained by STE and then refined in the DD domain through cross-correlation operation. However, the adopted TF domain model neglects ICI, which can cause model mismatch and severe estimation error floors in practical high-Doppler scenarios. In contrast, our previous work~\cite{nie2025novel} developed a novel CDCE for OFDM system that accounts for ICI and significantly improves estimation accuracy. In particular, the OFDM pilot sequence and its received signal are transformed into the DD domain, where DD components are coarsely estimated through two-dimensional twisted convolution. The TF domain channel is then reconstructed by solving a least-squares problem using either sparse recovery or pseudo-inverse operations, depending on the dimension of the resulting dictionary matrix.

In this paper, we propose a neural network (NN)-based DD-assisted channel estimation method for OFDM systems under high-Doppler channels. We first derive the OFDM input-output relation in the presence of fractional delay and Doppler, and then formulate channel estimation as a DD recovery problem. To address this problem, we develop an NN-based estimator that exploits nonlinear representation capability to estimate the channel under the pilot-data interference and ICI at the receiver. Compared with conventional compressed sensing methods, such as the least absolute shrinkage and selection operator (LASSO) \cite{gaudio2021otfs}, the proposed NN-based estimator can learn a nonlinear mapping from contaminated observations to channel parameters. Numerical results show that the proposed method significantly improves the normalized mean square error (NMSE) performance compared with conventional OFDM channel estimation schemes.

\emph{Notation:}
The superscripts $(\cdot)^{\rm{H}}$, $(\cdot)^{*}$ and $(\cdot)^{\top}$ denote the Hermitian transpose, conjugate and transpose operation, respectively; 
${\rm Tr}\{\cdot\}$ and ${\rm{vec}}\left( \cdot \right)$ denote the trace and the vectorization operation of a matrix; 
${{\bf{F}}_N}$ denotes the normalized discrete Fourier transform (DFT) matrix of size $N\times N$;
${\bf I}_M$ represents the $M\times M$ identity matrix;
``$ \otimes $" denotes the Kronecker product operator. 
$\mathbb{E}[\cdot]$ denotes the expectation operation;
$\mathbf{X}=\mathrm{diag}\left(\mathbf{x}\right)$ denotes a diagonal matrix $\mathbf{X}$ whose diagonal entries are given by vector $\mathbf{x}$. $\operatorname{blkdiag}\{\mathbf{A}_{1}, \mathbf{A}_{2}, \cdots \mathbf{A}_{N}\}$ stands for the block diagonal matrix with $\mathbf{A}_{1}, \mathbf{A}_{2}, \cdots \mathbf{A}_{N}$ being its diaognal.


\section{OFDM Transmissions over Doubly-Selective Channel}

Without loss of generality, let us consider an OFDM-based multi-carrier transmission system. Specifically, let $M$ be the number of subcarriers with subcarrier spacing $\Delta f$, and let $N$ be the number of OFDM symbols with each of duration $T$. The system operates under the critical sampling condition such that $\Delta f\cdot T=1 $. The transmitted symbols are represented in the TF domain by the matrix $\mathbf{X}_\mathrm{TF}\in\mathbb{C}^{M\times N}$. The corresponding time domain representation $\mathbf{S}\in\mathbb{C}^{M\times N}$ is then obtained by applying an inverse discrete Fourier transform (IDFT) along the frequency dimension as
\begin{align}
    \mathbf{S}=\mathbf{F}_{M}^{\mathrm{H}}\mathbf{X}_\mathrm{TF}.
\end{align}
By appending the CP of length $L_{\rm CP}$ to each OFDM symbol, we have
\begin{align}
    \tilde{\mathbf{S}}=\mathbf{A}_{\mathrm{CP}}\mathbf{S}=\mathbf{A}_{\mathrm{CP}}\mathbf{F}_{M}^{\rm H}\mathbf{X}_\mathrm{TF},
\end{align}
which can be written in vector form as
\begin{align}
    \tilde{\mathbf{s}}=(\mathbf{I}_N\otimes\mathbf{A}_{\mathrm{CP}})\mathbf{s}=(\mathbf{I}_N\otimes\mathbf{A}_{\mathrm{CP}}\mathbf{F}_{M}^{\rm H})\mathbf{x}_\mathrm{TF}.
\end{align}
where $\mathbf{s}\in\mathbb{C}^{MN\times 1}$ and $\tilde{\mathbf{s}}\in\mathbb{C}^{\tilde{M}N\times 1}$ denote the time domain transmit signal before and after inserting CP, and $\tilde{M}=M+L_{\rm CP}$. Moreover, $\mathbf{A}_{\mathrm{CP}}\triangleq[\mathbf{G}_{\mathrm{CP}},\mathbf{I}_{M}]^{\top}\in\mathbb{R}^{\tilde{M}\times M}$ is the OFDM CP addition matrix, where $\mathbf{G}_{\mathrm{CP}}$ of size $M\times L_{\mathrm{CP}}$ includes the last $L_{\mathrm{CP}}$ columns of the identity matrix $\mathbf{I}_{M}$. Subsequently, the continuous time domain transmit signal can be obtained by applying a pulse shaping $p(t)$, yielding
\begin{align}
    \tilde{s}(t)=\sum_{\tilde{n}=0}^{\tilde{M}N-1}{{s}}[\tilde{n}] p(t-\tilde{n}T_{\mathrm{s}}),
\end{align}
where ${{s}}[\tilde{n}]$ denotes the $\tilde{n}$-th element of $\tilde{\mathbf{s}}$ and $T_{\mathrm{s}}$ is the time domain sampling period satisfying $T_{\mathrm{s}}=\frac{T}{M}$. Here, we focus on a time-varying channel, whose sparse representation in the DD domain is given by
\begin{align}
    h(\tau,\nu)=\sum_{p=1}^{P}h_p\delta(\tau-\tau_p)\delta(\nu-\nu_p),
\end{align}
where $h_{p}$, $\tau_{p}=l_{p}\frac{1}{M\Delta f}$, and $\nu_{p}=k_{p}\frac{1}{NT}$ denote the channel coefficient, delay, and Doppler shifts of the $p$-th path, respectively. Note that $l_{p}$ and $k_{p}$ are not necessarily integers.

After experiencing the time-varying channel, the continuous time domain receive signal $\tilde{r}(t)$ can be expressed by
\begin{align}
    \tilde{r}(t)=\sum_{p=1}^{P} h_p \tilde{s}(t-\tau_p)e^{\mathrm{j} 2\pi\nu_p(t-\tau_p)}+w(t),
\end{align}
where $w(t)$ is the complex additive white Gaussian noise (AWGN) process with zero mean and one-sided power spectral density (PSD) $N_0$. By adopting the matched-filtering to the received signal, we obtain its discrete-time representation $\tilde{\mathbf{r}}\in\mathbb{C}^{\tilde{M}N\times 1}$, whose $\tilde{m}$-th element can be expressed as
\begin{align}
     \tilde{r}[\tilde{m}] &=  \int_{-\infty}^{\infty} \tilde{r}(t) p^*(t-\tilde{m}T_{\mathrm{s}}) \mathrm{d}t+w[\tilde{m}],\nonumber\\
    &=\sum_{\tilde{m}=0}^{\tilde{M}N-1}{{s}}[\tilde{n}]  g[\tilde{m},\tilde{n}]+w[\tilde{m}].\label{g_first}
\end{align}
Here, we define the effective time domain channel as 
\begin{align}
    g[\tilde{m},\tilde{n}] \triangleq\sum_{p=1}^{P} h_p e^{\mathrm{j} 2\pi \tilde{n} \nu_p T_{\mathrm{s}}} A^*((\tilde{n}-\tilde{m})T_{\mathrm{s}}+\tau_p,\nu_p),\label{g_element}
\end{align}
where $A^*((\tilde{n}-\tilde{m})T_{\mathrm{s}}+\tau_p,\nu_p)$ denotes the ambiguity function of pulse $p(t)$ with respect to delay $\tau_p$ and Doppler shift $\nu_p$, whose definition is given by
\begin{align}
    A(\tau_p,\nu_p)&\triangleq\int_{-\infty}^{\infty} p(t) p^*(t-\tau_p) e^{-j2\pi\nu_p(t-\tau_p)} \mathrm{d}t.
\end{align}
By stacking all received symbols into a vector and removing the CP, we obtain the following compact representation of the received signal:
\begin{align}
    \mathbf{r}&=(\mathbf{I}_{\mathrm{N}}\otimes\mathbf{R}_{\mathrm{CP}})\tilde{\mathbf{r}}=(\mathbf{I}_{\mathrm{N}}\otimes\mathbf{R}_{\mathrm{CP}})\sum_{p=1}^{P}\tilde{\mathbf{G}}_p \tilde{\mathbf{s}}+\mathbf{w}\nonumber\\
    &=(\mathbf{I}_{\mathrm{N}}\otimes\mathbf{R}_{\mathrm{CP}})\tilde{\mathbf{G}}_{\mathrm{T}}(\mathbf{I}_N\otimes\mathbf{A}_{\mathrm{CP}})\mathbf{s}+\mathbf{w}\nonumber\\
    &=\mathbf{G}_{\mathrm{T}}\mathbf{s}+\mathbf{w},\label{time domain IO}
\end{align}
where $\tilde{\mathbf{G}}_p$ represents the $p$-th resolvable path component of the time domain channel matrix $\tilde{\mathbf{G}}_{\mathrm{T}}$ before CP removal and the $(\tilde{m},\tilde{n})$-th element of $\tilde{\mathbf{G}}_{\mathrm{T}}$ is given in \eqref{g_element}. In addition, $\mathbf{R}_{\mathrm{CP}}\triangleq \left[\boldsymbol{0}_{M\times L_{\mathrm{CP}}}, \mathbf{I}_{M}\right]$ is the CP removal matrix, and $\mathbf{G}_{\mathrm{T}}$ denotes the equivalent time domain channel matrix after CP removal. Note that in \eqref{time domain IO} and what follows, we use $\mathbf{w}$ to denote the noise vector since it follows the same distribution.

Furthermore, the TF domain receive signal is obtained by applying a discrete Fourier transform (DFT) to $\mathbf{r}$, yielding
\begin{align}
    \mathbf{y}_\mathrm{TF}
    &=(\mathbf{I}_{{N}}\otimes\mathbf{F}_{M}) \mathbf{G}_{\mathrm{T}}  (\mathbf{I}_N\otimes\mathbf{F}_{M}^{\rm H})\mathbf{x}_\mathrm{TF}+\mathbf{w}, \\
    &=\mathbf{H}_\mathrm{TF}\mathbf{x}_\mathrm{TF}+\mathbf{w},\label{IO_TF}
\end{align}
where $\mathbf{H}_\mathrm{TF}\in\mathbb{C}^{MN\times MN}$ represents the TF domain channel matrix. Note that the channel matrix $\mathbf{H}_\mathrm{TF}$ exhibits a block diagonal structure, when the CP length is larger than the channel delay spread, i.e.,
\begin{align}\label{H_TF}
\mathbf{H}_\mathrm{TF} = \mathrm{blkdiag} \left\{  \mathbf{H}[1],\mathbf{H}[2], \cdots, \mathbf{H}[N] \right\},
\end{align}
where the diagonal elements of $\mathbf{H}[i]\in\mathbb{C}^{M\times M},1\le i\le N,$ represent the channel for the $i$-th OFDM symbol, while the off-diagonal elements account for the ICI.

\section{DD-Assisted Channel Estimation}

In this section, we develop a DD-assisted channel estimation framework based on the TF domain input-output relation (IOR) derived in the previous section. Unlike conventional OFDM channel estimation, which typically estimates only the diagonal effective channel coefficients at pilot positions by STE, the proposed framework further exploits the underlying DD parameters to enhance the accuracy of TF domain channel estimation.

To this end, we first rewrite the IOR in \eqref{IO_TF} by discretizing the delay and Doppler into a finite grid. Specifically, we define a DD grid $\Xi$, whose resolutions are determined by the delay and Doppler grid spacings. Let $Q$  denote the total number of grid points, given by the product of the Doppler and delay bins. Each grid point is denoted by $\xi_i=(\tau_i,\nu_i)\in\Xi, i = 0, \dots, Q-1$, where $\tau_i$ and $\nu_i$ are the corresponding delay and Doppler values, respectively. Then, we define the corresponding TF domain channel component $\check{\mathbf{H}}_{\xi_i}$ as
\begin{align}
    \check{\mathbf{H}}_{\xi_i}=(\mathbf{I}_{{N}}\otimes\mathbf{F}_{M}\mathbf{R}_{\mathrm{CP}}) \check{\mathbf{G}}_{\mathrm{T},\xi_i}  (\mathbf{I}_N\otimes\mathbf{A}_{\mathrm{CP}}\mathbf{F}_{M}^{\rm H}),
\end{align}
where $\check{\mathbf{G}}_{\mathrm{T},\xi_i}$ denotes the time domain channel component associated with the DD grid point $\xi_i$, whose $(\tilde{m},\tilde{n})$ element is given by
\begin{align}
g_{\xi_i}[\tilde{m},\tilde{n}]
=
e^{j2\pi \tilde{n}\nu_i T_{\mathrm{s}}}
A^*((\tilde{n}-\tilde{m})T_{\mathrm{s}}+\tau_i,\nu_i).
\end{align}
By vectorizing the matrices $\check{\mathbf{H}}_{\xi_i}$ for all grid points $\xi_i\in\Gamma$ and concatenating the resulting vectors, we construct a general dictionary matrix as
\begin{align}
    \mathbf{D}=\left[ \operatorname{vec}\left(\check{\mathbf{H}}_{\xi_0}\right),\dots, \operatorname{vec}\left(\check{\mathbf{H}}_{\xi_{Q-1}}\right)\right],
\end{align}
of dimension $(MN)^2\times Q$. We further define ${\mathbf{h}}\in\mathbb{C}^{Q\times 1}$ as the vector collecting the channel gains associated with the discrete DD components on $\xi_i\in\Xi$. Therefore, the vectorized TF domain channel matrix can be expressed as
\begin{align}
    \operatorname{vec}\left(\mathbf{H}_{\rm TF}\right) = \mathbf{D}{\mathbf{h}}.
\end{align}
Substituting this representation into the TF domain IOR in \eqref{IO_TF}, we obtain
\begin{align}
    \mathbf{y}_{\rm TF}=\left(\mathbf{x}_{\rm TF}^{\top}\otimes \mathbf{I}_{MN}\right)\mathbf{D}{\mathbf{h}}+\mathbf{w}=\mathbf{A}{\mathbf{h}}+\mathbf{w},\label{IO_TF_Ah}
\end{align}
where $\mathbf{A}=\left(\mathbf{x}_{\rm TF}^{\top}\otimes \mathbf{I}_{MN}\right)\mathbf{D}$ denotes the effective sensing matrix determined by the transmitted TF domain signal and the dictionary matrix.

It is worth noting that $\mathbf{h}$ is commonly estimated using sparse recovery techniques, which typically require constructing a sufficiently high-resolution dictionary matrix $\mathbf{D}$, such that the DD channel can be well approximated by a sparse vector $\mathbf{h}$. Representative sparse recovery solvers include LASSO, orthogonal approximate message passing (OAMP), and sparse Bayesian learning (SBL). However, these model-based approaches may suffer from high computational complexity and performance degradation when the true delay and Doppler values are off the predefined grid. To address these limitations, we next introduce a NN-based estimation approach that learns to recover the \emph{equivalent} DD domain channel from the TF domain observations. It should be emphasized that, although the physical DD domain channel is typically sparse, we do not impose sparsity on $\mathbf{h}$ in the proposed framework. Instead, $\mathbf{h}$ is interpreted as an \emph{equivalent} DD domain channel vector defined on a grid with limited DD resolution. As a result, off-grid delay and Doppler components may lead to energy leakage across neighboring DD grid points, making $\mathbf{h}$ generally non-sparse.
 

We now present the NN-based channel estimation scheme for estimating the equivalent DD domain vector $\mathbf{h}$, denoted by $\hat{\mathbf{h}}$. To facilitate NN processing, we first convert the complex-valued variables in \eqref{IO_TF_Ah} into their equivalent real-valued representations as
\begin{align}
\bar{\mathbf{y}}&=\left[\Re\{\mathbf{y}_{\rm TF}\}^{\top},\Im\{\mathbf{y}_{\rm TF}\}^{\top}\right]^{\top}\in\mathbb{R}^{2MN},\\
    \bar{\mathbf{h}}&=\left[\Re\{\mathbf{h}\}^{\top},\Im\{\mathbf{h}\}^{\top}\right]^{\top}\in\mathbb{R}^{2Q}, \\
    \bar{\mathbf{w}}&=\left[\Re\{\mathbf{w}\}^{\top},\Im\{\mathbf{w}\}^{\top}\right]^{\top}\in\mathbb{R}^{2MN},  
\end{align}
The complex-valued sensing matrix and dictionary matrix are rewritten in real-valued form as
\begin{align}
    \mathbf{\bar{A}}=
    \begin{bmatrix}
        \Re\{\mathbf{A}\} & -\Im\{\mathbf{A}\}\\
        \Im\{\mathbf{A}\} & \Re\{\mathbf{A}\}
    \end{bmatrix}\in\mathbb{R}^{2MN\times2Q},
\end{align}
and
\begin{align}
    \mathbf{\bar{D}}=
    \begin{bmatrix}
        \Re\{\mathbf{D}\} & -\Im\{\mathbf{D}\}\\
        \Im\{\mathbf{D}\} & \Re\{\mathbf{D}\}
    \end{bmatrix}\in\mathbb{R}^{2(MN)^2\times 2Q},
\end{align}
respectively.
Accordingly, the input-output relation in \eqref{IO_TF_Ah} and the vectorized TF domain channel can be rewritten, respectively, as
\begin{align}
    \bar{\mathbf{y}}=\mathbf{\bar{A}}\bar{\mathbf{h}}+\bar{\mathbf{w}},\quad
    \bar{\mathbf{g}}=\mathbf{\bar{D}}\bar{\mathbf{h}},
\end{align}
where $\bar{\mathbf{g}}=\left[\Re\{\operatorname{vec}(\mathbf{H}_{\rm TF})\}^{\top},\Im\{\operatorname{vec}(\mathbf{H}_{\rm TF})\}^{\top}\right]^{\top}$. In the proposed design, the NN takes the pair $(\mathbf{\bar{A}},\bar{\mathbf{y}})$ as input and outputs an estimate of the equivalent DD domain channel vector $\hat{{\mathbf{h}}}$, which is then used to reconstruct the TF domain channel as $\hat{{\mathbf{g}}}=\mathbf{\bar{D}}\hat{{\mathbf{h}}}$.

Following \cite{11113418}, the proposed NN-based channel estimator consists of three main components: feature representation, feature processing, and feature exchange modules.
We denote the observation feature and estimation feature at iteration $l$ by $\mathbf{F}_{\rm y}^{(l)}\in \mathbb{R}^{2MN\times d}$ and $\mathbf{F}_{\rm h}^{(l)}\in\mathbb{R}^{2Q\times d}$, respectively, where $d$ is the feature dimension and $l=0,\ldots,L$ denotes the feature update index. The observation feature is initialized by repeating the received signal over the feature dimension, i.e., $\mathbf{F}_{\rm y}^{(0)}=\bar{\mathbf{y}}\textbf{1}_d^{\top}$, where $\textbf{1}_d\in\mathbb{R}^d$ is an all-one vector. 

The feature processing modules are implemented by shared multi-layer perceptrons (MLPs). Specifically, $\Phi^{(l)}(\cdot)$ denotes the feature processor for updating the observation feature $\mathbf{F}_{\rm y}$, while $\Psi^{(l)}(\cdot)$ denotes the feature processor for updating the channel-estimation feature $\mathbf{F}_{\rm h}$. For $l<L$, both MLPs map each $d$-dimensional feature vector to another $d$-dimensional feature vector. At the final iteration, $\Psi^{(L)}(\cdot)$ maps each $d$-dimensional channel feature to a scalar, thereby producing the estimated real-valued DD domain channel vector $\hat{{\mathbf{h}}}\in\mathbb{R}^{2Q}$.

The feature exchange module maps information between the two feature representations through the sensing matrix $\bar{\mathbf{A}}$. In particular, the observation feature $\mathbf{F}_{\rm y}$ is projected to the estimation feature $\mathbf{F}_{\rm h}$ by $\bar{\mathbf{A}}^{\top}$, i.e., $\mathbf{F}_{\rm h}^{(l)}=\bar{\mathbf{A}}^{\top}\mathbf{F}_{\rm y}^{(l)}$, while the estimation feature is projected back to the observation feature by $\bar{\mathbf{A}}$, i.e., $\mathbf{F}_{\rm y}^{(l)}=\bar{\mathbf{A}}\mathbf{F}_{\rm h}^{(l)}$.

Starting from $\mathbf{F}_{\rm y}^{(0)}$, the $l$-th iteration proceeds as follows. First, the observation feature is updated by $\Phi^{(l)}(\cdot)$ and then projected to the estimation domain as $\mathbf{F}_{\rm h}^{(l)}
=
\bar{\mathbf{A}}^{\top}
\Phi^{(l)}\left(\mathbf{F}_{\rm y}^{(l-1)}\right)$. For $l>1$, a weighted skip connection is added to reuse the previous estimation feature as ${\mathbf{F}}_{\rm h}^{(l)}
= {\mathbf{F}}_{\rm h}^{(l)} + \alpha \mathbf{F}_{\rm h}^{(l-1)}$, where $\alpha$ is a skip-connection weight. The estimation feature is then updated as $\mathbf{F}_{\rm h}^{(l)}
=\Psi^{(l)}\left(\mathbf{F}_{\rm h}^{(l)}\right)$. Next, $\mathbf{F}_{\rm h}^{(l)}$ is projected back to the observation domain, and the repeated observation feature is subtracted to form a residual-type observation feature for the next iteration as $\mathbf{F}_{\rm y}^{(l)}
=\bar{\mathbf{A}}\mathbf{F}_{\rm h}^{(l)}-\bar{\mathbf{y}}\mathbf{1}_{d}^{\top}.$ After $T$ iterations, the final channel feature is converted into the estimated DD domain channel vector by $\hat{{\mathbf{h}}} = \Psi^{(L)}\left({\mathbf{F}}_{\rm h}^{(L)}\right).$ The corresponding TF domain channel vector is then reconstructed as $\hat{{\mathbf{g}}}=\bar{\mathbf{D}}\hat{{\mathbf{h}}}$. The overall procedure is summarized in Algorithm~\ref{alg:nn_ce}.

The NN parameters are trained in a supervised manner by minimizing the TF domain channel reconstruction error. Specifically, the loss function is defined as the mean-square error (MSE) between the true and reconstructed TF domain channel vectors:
\begin{align}
    \mathcal{L}
    =
    \rm{MSE}(\bar{\mathbf{D}}\hat{{\mathbf{h}}}-\bar{\mathbf{g}}).
\end{align} 
The parameters of the feature processors $\Phi^{(l)}(\cdot)$ and $\Psi^{(l)}(\cdot)$ are then optimized via backpropagation.

\begin{algorithm}[t]
\caption{NN-based Channel Estimation}
\label{alg:nn_ce}
\begin{algorithmic}[1]
\State \textbf{Input:} $\bar{\mathbf{A}}$, $\bar{\mathbf{D}}$, and $\bar{\mathbf{y}}$
\State Initialize $\mathbf{F}_{\rm y}^{(0)}=\bar{\mathbf{y}}\mathbf{1}_{d}^{\top}$ 
\For{$l=1,\ldots,L$}
    \State $\mathbf{F}_{\rm y}^{(l)}=\Phi^{(l)}(\mathbf{F}_{\rm y}^{(l-1)})$
    \State $\mathbf{F}_{\rm h}^{(l)}=\bar{\mathbf{A}}^{\top}\mathbf{F}_{\rm y}^{(l)}$
    \If{$l>1$}
        \State $\mathbf{F}_{\rm h}^{(l)}=\mathbf{F}_{\rm h}^{(l)}+\alpha\mathbf{F}_{\rm h}^{(l-1)}$
    \EndIf
    \State $\mathbf{F}_{\rm h}^{(l)}=\Psi^{(l)}(\mathbf{F}_{\rm h}^{(l)})$
  
    \State $\mathbf{F}_{\rm y}^{(l)}=\bar{\mathbf{A}}\mathbf{F}_{\rm h}^{(l)}$
    \State $\mathbf{F}_{\rm y}^{(l)}=\mathbf{F}_{\rm y}^{(l)}-\bar{\mathbf{y}}\mathbf{1}_{d}^{\top}$

\EndFor 
\State \textbf{Output:} $\hat{{\mathbf{h}}}=\mathbf{F}_{\rm h}^{(L)}$, $\hat{{\mathbf{g}}}=\bar{\mathbf{D}}\hat{{\mathbf{h}}}$
\end{algorithmic}
\end{algorithm}

\section{Numerical Results}
\subsection{System Settings}
\begin{figure}
    \centering
    \includegraphics[scale=0.5]{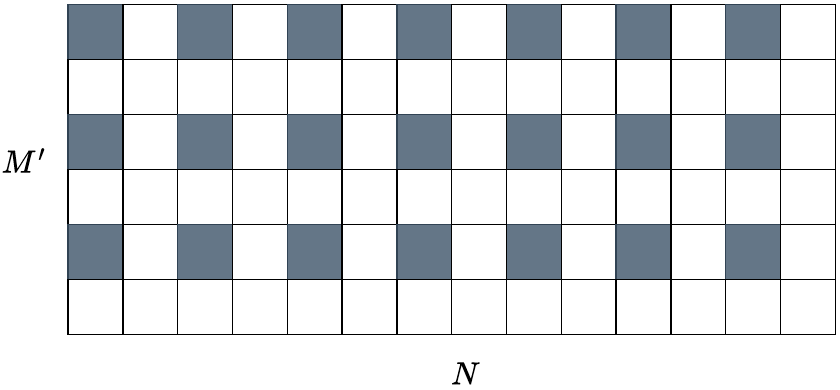}
    \caption{Pilot arrangement in the TF domain in $\mathbf{X}_{\mathrm{TF}}$.}
    \label{fig:pilots}
\end{figure}

The system is configured with $M=N=8$. The number of channel paths is set to $P=3$. For each path, the delay and Doppler indices are independently and uniformly sampled from $[0,l_{\max}]$ and $[-k_{\max},k_{\max}]$, respectively, where $l_{\max}=3$ and $k_{\max}=2$ denote the maximum delay and Doppler indices, respectively. Note that the delay and Doppler indices are not restricted to integer values. We adopt a lattice pilot structure, where pilots are placed along both time and frequency dimensions, making it suitable for channel estimation in doubly selective environments. The TF domain pilots are placed in $\mathbf{X}_{\mathrm{TF}}$ and set to all ones, as shown in Fig. \ref{fig:pilots}. The data symbols are independently drawn from quadrature phase-shift keying (QPSK) modulation with unit power. The SNR is defined with respect to the received signal power. For each channel realization, the noiseless received TF domain signal is given by $\mathbf{y}_{0}=\mathbf{H}_{\mathrm{TF}}\mathbf{x}_{\rm TF}$, and the average received signal power is computed as $\mathcal{P}_{y}=\frac{1}{MN}||\mathbf{y}_{0}||^2_2$. Accordingly, the SNR is defined as $\mathrm{SNR} = \frac{\mathcal{P}_{y}}{N_{0}}$. The normalized MSE (NMSE) is defined as $\operatorname{NMSE} = \frac{||\hat{\mathbf{H}}_{\rm TF}-\mathbf{H}_{\rm TF}||^2_2}{||\mathbf{H}_{\rm TF}||^2_2}$.

To evaluate the performance of the proposed channel estimation method, several conventional OFDM channel estimation schemes are considered as benchmarks. Specifically, STE is first applied at the pilot locations to obtain the channel state information (CSI). The CSI at non-pilot locations is then recovered by linear interpolation along both time and frequency dimensions, using the nearest available pilot-based estimates. The STE based on least square (LS) and linear minimum MSE (LMMSE) estimators are adopted, referred to as ``ST-LS'' and ``ST-LMMSE'', respectively. These estimators are expressed as
\begin{align}
   \operatorname{vec}(\mathbf{H}_{\rm TF})_{\mathrm{ST-LS}}&=\mathbf{y}_{\mathrm{TF}}\oslash\mathbf{x}_{\mathrm{TF}},\\
   \operatorname{vec}(\mathbf{H}_{\rm TF})_{\mathrm{ST-LMMSE}}&=\frac{1}{1+\frac{1}{\mathrm{SNR}}}\odot\operatorname{vec}(\mathbf{H}_{\rm TF})_{\mathrm{ST-LS}},
\end{align}
where $\oslash$ and $\odot$ denote the element-wise division and Hadamard product, respectively.

\subsection{Training Settings}
Both feature processors $\Phi^{(t)}(\cdot)$ and $\Psi^{(t)}(\cdot)$ are implemented as MLPs comprising an input layer, a hidden layer of dimension $d_{h}$, and an output layer. At each iteration, the MLPs map input feature matrices in $\mathbb{R}^{2MN\times d}$ to output feature matrices of the same dimension. The only exception is $\Psi^{(T)}(\cdot)$ in the final iteration, whose output layer produces a scalar coefficient. In the simulations, we set $d=12$, $d_{\rm h}=64$, $L=10$, and $\alpha=0.5$ for skip connection. 
The training data are generated according to the OFDM input-output model in \eqref{IO_TF_Ah}. Specifically, we generate $50000$ training samples for each SNR value in $\{10,15,20,25\}$ dB.
During training, an SNR value is randomly selected from this set, and each mini-batch is formed by randomly sampling examples generated at the selected SNR. The network is trained using the Adam optimizer with a learning rate of $10^{-4}$ and a weight decay of $10^{-4}$. Unless otherwise specified, the mini-batch size and number of training epochs are set to $256$ and $200$, respectively.

\subsection{Simulation results}

\begin{figure}
    \centering
    \includegraphics[scale=0.5]{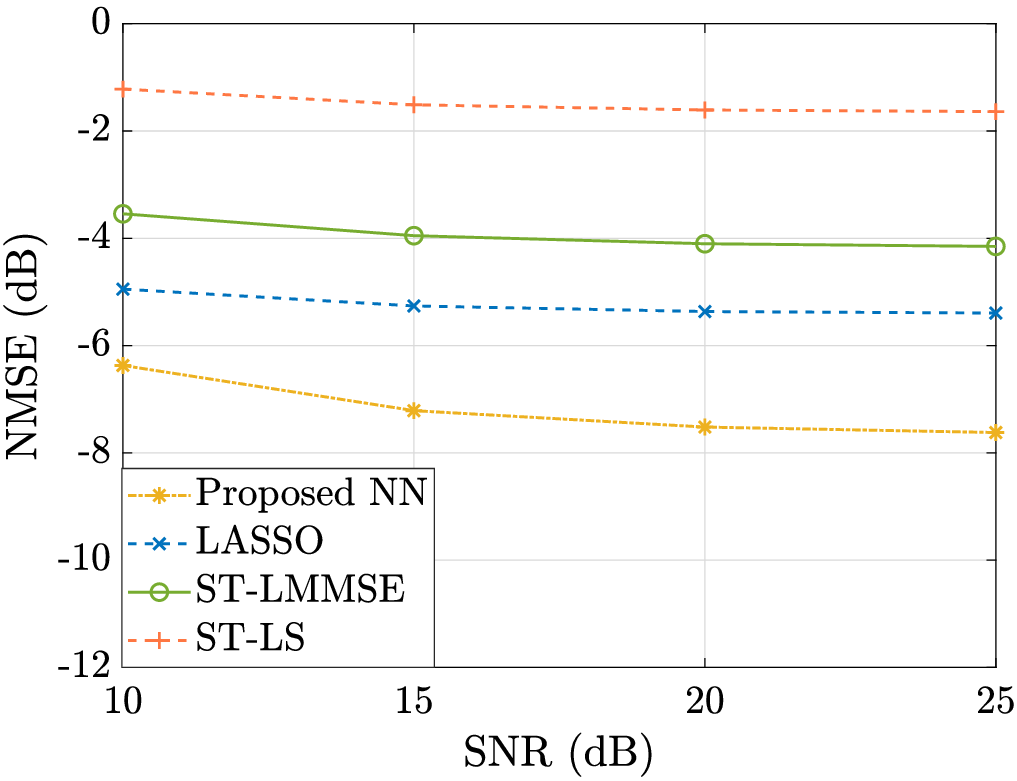}
    \caption{NMSE comparison for different estimation schemes.}
    \label{fig:NMSE}
\end{figure}

As shown in Fig. \ref{fig:NMSE}, conventional OFDM channel estimation schemes suffer from severe NMSE degradation in high-mobility environments. This is mainly because strong Doppler shifts introduce significant ICI, whereas STE assumes that the TF domain input-output relationship can be approximated by an element-wise product. As a result, STE neglects the off-diagonal interference components in the TF domain channel matrix during the estimation stage. Among the STE-based methods, ST-LMMSE outperforms ST-LS because it further exploits the second-order statistics of the channel. In contrast, DD-assisted schemes achieve better NMSE performance by estimating the underlying DD channel parameters and reconstructing the corresponding TF domain channel. Specifically, the model in \eqref{IO_TF_Ah} can be formulated as a sparse recovery problem, where LASSO is employed to estimate the equivalent DD domain channel vector. The estimated DD domain channel is then used to construct the full TF domain channel matrix, thereby partially accounting for the ICI effect. However, the performance of LASSO may still be limited by the interference between pilot and data symbols at the receiver, as well as by possible model mismatch in the sparse representation. The proposed NN-based DD-assisted channel estimation method achieves the most robust performance among all considered schemes. This indicates that the NN can better exploit the structured DD domain information while mitigating the adverse effects of ICI and pilot-data interference.

\begin{figure}
    \centering
    \includegraphics[scale=0.5]{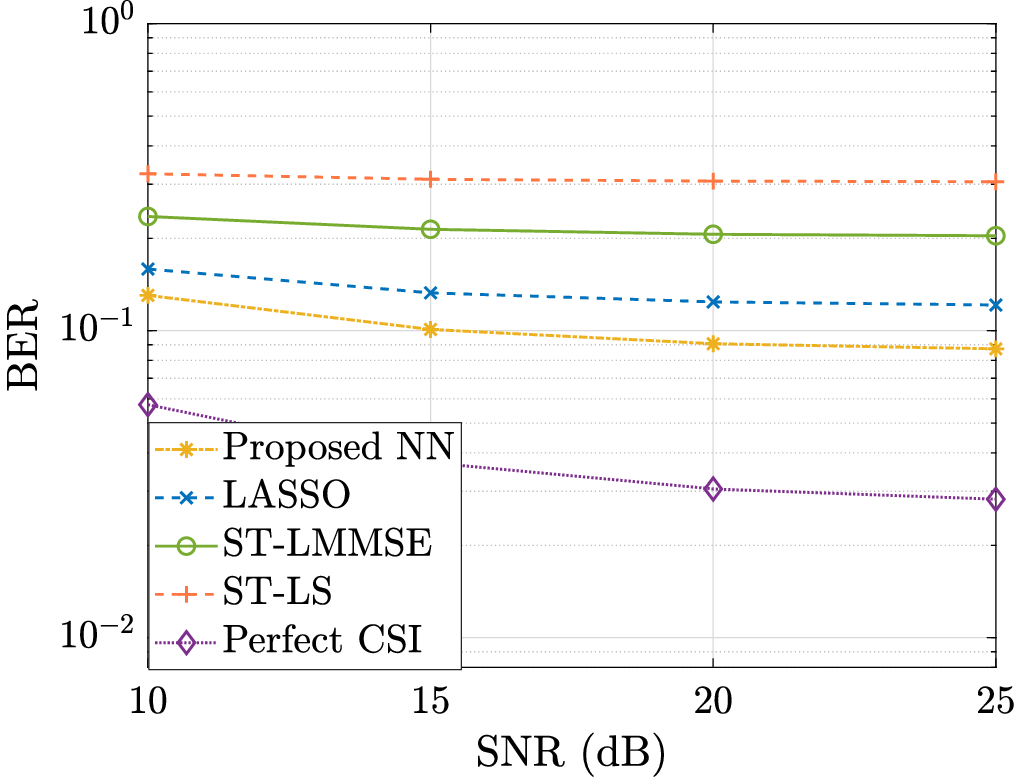}
    \caption{BER comparison for different estimation schemes.}
    \label{fig:BER}
\end{figure}

Moreover, the bit-error-rate (BER) performance is presented in Fig. \ref{fig:BER}. It can be observed that all schemes exhibit a trend consistent with their NMSE performance. Specifically, the conventional ST-LS and ST-LMMSE schemes achieve the worst BER performance, since they fail to properly account for the ICI effect in high-mobility scenarios. The LASSO-based DD-assisted scheme improves the BER by exploiting the sparse structure of the DD domain channel. In comparison, the proposed NN-based DD-assisted scheme achieves the best BER performance among all practical channel estimation schemes, demonstrating its superior robustness against ICI and pilot-data interference. Nevertheless, a noticeable performance gap still remains between the proposed scheme and the perfect-CSI benchmark. This indicates that the channel estimation accuracy is still the dominant factor limiting the detection performance, and further improvement is possible by enhancing the DD domain parameter estimation.

\section{Conclusion}

In this paper, we proposed a NN-based DD-assisted channel estimation framework for OFDM systems in high-mobility environments. By deriving an ICI-aware input-output relation, channel estimation was formulated as the recovery of an equivalent DD domain channel representation, from which the TF domain channel matrix was reconstructed. Unlike conventional sparse recovery methods, the proposed framework does not require the equivalent DD domain channel vector to be strictly sparse and can therefore accommodate the leakage caused by fractional delay and Doppler shifts. Moreover, because the sensing matrix is constructed using only known pilot symbols, the resulting pilot-only model does not fully capture data-induced interference. To address this structured model mismatch, the adopted network iteratively exchanges observation- and channel-domain features through the sensing matrix to learn the recovery of the DD domain channel from contaminated pilot observations. Simulation results demonstrated that the proposed method achieves lower NMSE and BER than conventional single-tap LS and LMMSE estimators, as well as the LASSO-based DD-assisted estimator.


\bibliographystyle{IEEEtran}
\bibliography{ref}

@article{wei2021orthogonal,
          title={Orthogonal time-frequency space modulation: A promising next-generation waveform},
          author={Wei, Zhiqiang and Yuan, Weijie and Li, Shuangyang and Yuan, Jinhong and Bharatula, Ganesh and Hadani, Ronny and Hanzo, Lajos},
          journal={IEEE Trans. Wireless Commun.},
          volume={28},
          number={4},
          pages={136--144},
          year={Aug. 2021},
}

@article{raviteja2019embedded,
    title={Embedded pilot-aided channel estimation for {OTFS} in delay-{Doppler} channels},
    author={Raviteja, Patchava and Phan, Khoa T and Hong, Yi},
    journal={IEEE Trans. Veh. Technol.},
    volume={68},
    number={5},
    pages={4906--4917},
    year={May 2019},
    }

@article{gaudio2021otfs,
  title={{OTFS vs. OFDM} in the presence of sparsity: A fair comparison},
  author={Gaudio, Lorenzo and Colavolpe, Giulio and Caire, Giuseppe},
  journal={IEEE Trans. Wireless Commun.},
  volume={21},
  number={6},
  pages={4410--4423},
  year={Dec. 2021},
}

@inproceedings{hadani2017orthogonal,
    title={Orthogonal time frequency space modulation},
    author={Hadani, Ronny and Rakib, Shlomo and Tsatsanis, Michail and Monk, Anton and Goldsmith, Andrea J and Molisch, Andreas F and Calderbank, R},
    booktitle={Proc. IEEE WCNC},
    pages={1--6},
    year={Mar. 2017},
}

@misc{3GPP38211,
    author = {{3rd Generation Partnership Project (3GPP)}},
    title = {{3GPP TS 38.211: NR; Physical channels and modulation}},
    howpublished = {Online},
    year = {2024},
    note = {Available: \url{https://www.etsi.org/deliver/etsi_ts/138200_138299/138211/16.02.00_60/ts_138211v160200p.pdf}}
}

@article{hu2024cross,
  title={Cross-Domain Channel Estimation Based Serial Interference Cancellation in {NOMA-OTFS} System},
  author={Hu, Jiacheng and Bai, Zhiquan and Xu, Hao and Liu, Hongwu and Wang, Yingxun and Kwak, KyungSup},
  journal={IEEE Commun. Lett.},
  volume={28},
  number={7},
  pages={1668--1672},
  year={Jul. 2024},
}

@article{nie2024uplink,
  title={Uplink multi-user {OTFS}: Transmitter design based on statistical channel information},
  author={Nie, Mingcheng and Li, Shuangyang and Mishra, Deepak and Yuan, Jinhong and Ng, Derrick Wing Kwan},
  journal={IEEE Trans. on Commun.},
  volume={73},
  number={7},
  pages={4678--4696},
  year={Dec. 2024},
}

@inproceedings{nie2025novel,
  title={A novel cross-domain channel estimation scheme for {OFDM}},
  author={Nie, Mingcheng and Chong, Ruoxi and Li, Shuangyang and Yuan, Weijie and Ng, Derrick Wing Kwan and Matthaiou, Michalis and Caire, Giuseppe and Li, Yonghui},
  booktitle={IEEE GLOBECOM Proceedings},
  year={2025},
}

@ARTICLE{11113418,
  author={Ye, Hao and Liang, Le},
  journal={IEEE Transactions on Signal Processing}, 
  title={On Purely Data-Driven Massive MIMO Detectors}, 
  year={2025},
  volume={73},
  number={},
  pages={3079-3093},
  doi={10.1109/TSP.2025.3594197}}

@INPROCEEDINGS{10225901,
  author={Chang, Hao and Kosasih, Alva and Hardjawana, Wibowo and Qu, Xinwei and Vucetic, Branka},
  booktitle={IEEE INFOCOM WKSHPS}, 
  title={Untrained Neural Network based Bayesian Detector for OTFS Modulation Systems}, 
  year={2023},
  volume={},
  number={},
  pages={1-6},
  doi={10.1109/INFOCOMWKSHPS57453.2023.10225901}}

@INPROCEEDINGS{10683123,
  author={Chang, Hao and Vucetic, Branka and Hardjawana, Wibowo},
  booktitle={IEEE Veh Technol Conf}, 
  title={Graph-Based Untrained Neural Network Detector for OTFS Systems}, 
  year={2024},
  volume={},
  number={},
  pages={1-6},
  doi={10.1109/VTC2024-Spring62846.2024.10683123}}

@ARTICLE{11153399,
  author={Chang, Hao and Pang, Gaoyang and Vucetic, Branka and Hardjawana, Wibowo},
  journal={IEEE Wireless Commun. Lett.}, 
  title={Neural OTFS Receiver Without Training Requirements}, 
  year={2025},
  volume={14},
  number={11},
  pages={3824-3828},
  doi={10.1109/LWC.2025.3606969}}

@ARTICLE{11373535,
  author={Nie, Mingcheng and Chong, Ruoxi and Li, Shuangyang and Farhang, Arman and Göttsch, Fabian and Ng, Derrick Wing Kwan and Matthaiou, Michail and Li, Yonghui},
  journal={IEEE Commun. Stand. Mag.}, 
  title={Toward Standardizing {OTFS}: A Candidate Waveform for Next-Generation Wireless Networks}, 
  year={Feb. 2026},
  volume={},
  number={},
  pages={1-12},
  doi={10.1109/MCOMSTD.2026.3657606}}

@inproceedings{nie2023improving,
  title={Improving channel estimation performance for uplink {OTFS} transmissions: Pilot design based on a posteriori {Cram{\'e}r-Rao} bound},
  author={Nie, Mingcheng and Li, Shuangyang and Mishra, Deepak},
  booktitle={Proc. IEEE Int. Conf. Commun. Workshops (ICC Workshops)},
  pages={301--306},
  year={2023},
}

@ARTICLE{10654761,
  author={Zhang, Xiaoqi and Liu, Chang and Yuan, Weijie and Zhang, J. Andrew and Ng, Derrick Wing Kwan},
  journal={IEEE Trans. Veh. Technol.}, 
  title={{Sparse Prior-Guided Deep Learning for OTFS Channel Estimation}}, 
  year={Aug. 2024},
  volume={73},
  number={12},
  pages={19913-19918},
  doi={10.1109/TVT.2024.3450012}}

@article{zhang2025deep,
  title={{Deep Learning-based OTFS Channel Estimation and Symbol Detection with Plug-and-Play Framework}},
  author={Zhang, Xiaoqi and Ni, Zhitong and Yuan, Weijie and Zhang, J Andrew and Quek, Tony QS},
  journal={IEEE Trans. on Commun.},
  year={Nov. 2025},
}

@ARTICLE{wong2026fairness,
  author  = {C. H. A. Wong and D. Mishra and M. Nie and A. Shafie and J. Yuan},
  title   = {Fairness-Aware {ODDM} Design for Secrecy Rate Maximization Among Untrusted Users},
  journal = {IEEE Wireless Commun. Lett.},
  volume  = {15},
  pages   = {2184--2188},
  year    = {Feb. 2026},
  doi     = {10.1109/LWC.2026.3668884}
}

\newpage

\vfill

\end{document}